\documentstyle[aaspp4,12pt,epsf]{article}
\def\lsim{\, \lower2truept\hbox{${< \atop\hbox{\raise4truept\hbox{$\sim$}}}$}\,}
\def\gsim{\, \lower2truept\hbox{${> \atop\hbox{\raise4truept\hbox{$\sim$}}}$}\,}
\def\nid{\noindent}
\def\oneskip{\vskip\baselineskip}
\def\sun{\odot}

\def\puncspace{\ifmmode\,\else{\ifcat.\C{\if.\C\else\if,\C\else\if?\C\else%
\if:\C\else\if;\C\else\if-\C\else\if)\C\else\if/\C\else\if]\C\else\if'\C%
\else\space\fi\fi\fi\fi\fi\fi\fi\fi\fi\fi}%
\else\if\empty\C\else\if\space\C\else\space\fi\fi\fi}\fi}
\def\SP{\let\\=\empty\futurelet\C\puncspace}
\def\micron{~$\mu$m\SP}

\def\etal{~~{\it et al.}\SP}

\begin{document}

\title{Emission Features and Source Counts of Galaxies in Mid-Infrared}
\author{Cong Xu, Perry B. Hacking, Fan Fang, David L. Shupe, Carol
J. Lonsdale, Nanyao Y. Lu, George Helou,}
\affil{Infrared Processing and Analysis Center, Jet Propulsion Laboratory, 
Caltech 100-22, Pasadena, CA 91125}
\author{Gordon J. Stacey}
\affil{Department of Astronomy, Cornell University, Ithaca, NY 14853}
\author{Matthew L.~N.~Ashby}
\affil{Smithsonian Astrophysical Observatory \\ Optical \& Infrared Astronomy
Division \\ 60 Garden Street MS 66, Cambridge, MA 02318}
\begin{abstract}
In this work we incorporate the newest ISO results on the mid-infrared
spectral-energy-distributions (MIR SEDs)
of galaxies into models for the number counts and 
redshift distributions of MIR surveys. A three-component model,
with empirically determined MIR SED templates of (1) 
a cirrus/PDR component (2)
a starburst component and (3) an AGN component, is developed for 
infrared (3--120\micron) SEDs of galaxies.
The model includes a complete IRAS 25\micron selected
sample of 1406 local galaxies ($z \leq 0.1$; Shupe et al. 1998a). 
Results based on these
1406 spectra show that the MIR emission features
cause significant effects on the redshift dependence of the K-corrections
for fluxes in the WIRE 25\micron
band and ISOCAM 15\micron band. This in turn will affect
deep counts and redshift distributions in these two bands, as shown by 
the predictions of two evolution models (a 
luminosity evolution model with $L\propto (1+z)^3$ and a
density evolution model with $\rho\propto (1+z)^4$).
The dips-and-bumps on curves of MIR
number counts, caused by the emission features, 
should be useful indicators of evolution mode. The strong emission
features at $\sim 6$--8\micron will help the detections
of relatively high redshift ($z\sim 2$) galaxies in MIR surveys. 
On the other hand, determinations of the evolutionary
rate based on the slope of source counts, and studies on the large scale
structures using the redshift distribution of MIR sources, will
have to treat the effects of the MIR emission features carefully.
We have also estimated a 15\micron local luminosity 
function from the
predicted 15\micron fluxes of the 1406 galaxies using the bivariate 
(15\micron vs. 25\micron luminosities) method. This luminosity
function will improve our understanding of the ISOCAM 15\micron surveys.

\end{abstract}

\keywords{galaxies: starburst -- Seyfert -- luminosity function; 
infrared: galaxies}

\section{Introduction}

Current research on cosmology is focused on the debate whether
the star formation history of the universe has a sharp peak in the
redshift range of $z \sim 1$ -- 2 (Madau\etal 1996), as derived from 
the Hubble Space Telescope deep surveys  
(Williams\etal 1996; Griffiths\etal 1994)
and from ground based redshift surveys 
(Lilly\etal 1995; Koo and Kron 1992), or the universe had been
forming stars in a relatively constant rate since an early galaxy formation
time ($z \sim 5$) until $z\sim 1$ (Franceschini\etal 1997;
Rowan-Robinson\etal 1997), as suggested by the cosmic far-infrared
(FIR) background
studies (Puget\etal 1996; Guiderdoni\etal 1997) and studies
on the metal abundance in Intra-Cluster Medium (Mushotzky and
Loewenstein 1997). 

The problem with the HST surveys and ground
based redshift surveys is the dust extinction,
which may hide much of the star formation in early universe from 
these UV/optical surveys. 
Dust extinction has little effect on mid-infrared 
(MIR, 4\micron --- 40\micron) 
surveys. Compared to the FIR ($>$40\micron) deep surveys,
the MIR  surveys also have the advantage of using larger detector arrays
and having better resolutions;
the latter is of particular importance because in these bands
the depth of a survey is ultimately constrained by 
the confusion limit. Therefore, MIR surveys will play important roles
in galaxy evolution studies. Indeed, currently
several deep MIR surveys are either
being carried out using ISO (Franceschini\etal 1996; Oliver\etal 1997), 
or soon to be launched such as WIRE (Hacking\etal 1996;
Schember\etal 1996) and SIRTF (Cruikshank and Werner 1997).

On the other hand, it was known before ISO that in the 
wavelength range between 3--20\micron there are several broad band
features (see Puget and L\'eger 1989 for a review), 
which may contribute substantially to the IRAS 12\micron
flux of galaxies (Gillett\etal 1975;  Phillips\etal 1984;
Giard\etal 1989; Helou\etal 1991). The best
candidates for the emitters of these features are some large molecules
($\sim$1~nm) of Polycyclic Aromatic Hydrocarbon (L\'eger and Puget
1984; Allamandola\etal 1985), 
and the features are often called PAH features. New ISO observations
(Lu\etal 1996, 1998; Vigroux\etal 1996; Boulade\etal 1996) show
further that these features are present in the MIR spectra
of most galaxies, and their equivalent widths can be as large as 
$\sim$10\micron, comparable to or even larger than 
typical band passes for MIR filters (e.g. the band pass
of the ISOCAM 15\micron filter is $\sim$6\micron). This indicates that when
any of these features moves in or out of the band pass of a MIR
survey, because of the redshifts of galaxies, very significant
K-corrections will occur. Interpretations of MIR survey counts have
to take this fact into account. 
A pre-ISO attempt on
this issue, based on the laboratory PAH spectra (instead of
the spectra of galaxies), can be found in Maffei (1994). 

It is the primary aim of this paper to develop a new, emiprical model 
for the MIR SED's of galaxies, and incorprate it into the model for
number counts of MIR surveys. The SED model is built upon
available MIR spectra of galaxies
taken from published ISO observations, and is applied to the 1406 galaxies
(with $z<0.1$) 
of the 25\micron sample of Shupe\etal (1998a), selected from the
IRAS Faint Point Source Catalog (Moshir\etal 1992). 
For each galaxy in
the sample a model spectrum is obtained, which has the MIR emission 
features and which
can reproduce the four observed IRAS fluxes. These 1406
spectra are used to
determine the K-corrections and their dispersions
in our model for predictions on
number counts and redshift distributions of MIR surveys.
The effects of the emission features on these predictions are
investigated. Detailed comparisons of the model results with some 
ISOCAM MIR surveys (Oliver\etal 1997; Clements\etal 1998) and
predictions for WIRE survey will be found in a separate paper (Hacking
et al. 1998). The computer code of the model presented here
is available upon request to CX.

The paper is arranged as follows: the SED model is developed in Section 2. 
A local luminosity function at 15\micron, calculated from
the predicted 15\micron fluxes of the 1406 galaxies, is presented in
section 3. In Section 4 
we present the model for source counts and
investigate the effects of the emission features on the MIR number
counts. Section 5 is devoted to the discussion and 
Section 6 to the conclusion.

\section{An empirical model for IR SEDs of galaxies}
\subsection{Properties of MIR SEDs of galaxies}
The following properties of MIR SEDs of galaxies can be found from
the literature:
\begin{enumerate}
\item The MIR SEDs of normal galaxies (without AGNs) 
in the $\sim$10\micron region are not smooth.
Indeed there has been evidence for this 
since the observations of Gillett\etal (1975) of the 8--13\micron
spectra of M~82 and NGC~253, the two nearby prototype starburst galaxies.
Later ground-based observations (Roche\etal 1991) and the IRAS
low-resolution spectral observations (Cohen and Klein 1989) show
further that the 11.3\micron feature and the 8\micron feature are
very common in the 8 -- 13\micron spectra of 
`HII' galaxies (i.e. starburst galaxies). Most recently, 
Lu\etal (1996, 1998) obtained MIR spectra (3--11.7\micron)
of a sample of late-type galaxies using ISO. These 
galaxies are part of a larger set of star-forming galaxies,
selected to cover a broad range of infrared and optical properties, 
in the ISO key project on normal galaxies (Helou\etal 1996).
 The spectra were taken with ISOPHT operated in
its PHT-S spectroscopic mode (Lemke\etal 1996) with a triangular
chopping on both sides of the galaxy on the sky.  
In the wavelength range covered by the observations, 3--11.7\micron, several
prominent broad band features (e.g. at 6.2\micron, 
7.7\micron, 8.6\micron and 11.3\micron) are clearly visible in almost
every spectrum in the sample. ISOCAM CVF spectrum of NGC~5195 (the
small companion of M51) shows further
that beyond the PHT-S long wavelength boundary
there is still another broad feature at $\sim$12.7\micron
(close to the [NII] line at 12.8\micron; Boulade\etal 1996). 
All of these features are presumably
emission bands of PAH molecules
corresponding to different C-C and C-H modes (L\'eger and Puget 1984;
Allamandola\etal 1985).
\item Viewing through the gallery of ISO MIR spectra of
galaxies (Lu\etal 1996, 1998; Boulade\etal 1996; Vigroux\etal 1996;
Metcalfe\etal 1996; Acosta-Pulido\etal 1996), of Galactic
star formation regions and reflection nebulae
(Verstraete\etal 1996; Cesarsky\etal 1996a,
1996b; Boulanger\etal 1996), and of Galactic cirrus (Mattila\etal 1996),
it appears that in general, the PAH features and the underlying continuum
in the MIR band correlate tightly, indicating that they may have
the same origin. Except for the HII regions and violent starburst regions
in interacting galaxies, the shape of the MIR spectra of galaxies,
of different Galactic nebulae and of Galactic cirrus are remarkably similar.
However, in the core of Galactic HII regions (Cesarsky\etal1996b;
Verstraete\etal 1996) and in a violent starburst region in
Arp~244 (the famous Antennae which is a local example of galaxy merger),
another component of the continuum which increases sharply with
wavelength appears (Cesarsky\etal 1996b; Vigroux\etal 1996; 
Verstraete\etal 1996). In the framework of the dust model of D\'esert
et al. (1990), this latter component
is likely to be due to the 3-dimensional very small grains heated by
the combined heating of multiple photons (Boulanger\etal 1996). 
The characteristic temperature of
this component is usually too low for it to be seen at wavelengths
$<$20\micron when the heating radiation intensity is 
below $\sim 10^4$ times the local ISRF (Cesarsky\etal 1996b). Noticeably,
this component affects mostly the MIR emission at
$\lambda >$12\micron and not so much for the emission
at shorter wavelengths as shown by the ISO PHT-S spectra 
(3 -- 11.7\micron) of  
galaxies in the sample of Lu\etal (1998) which includes
both quiescent and starburst galaxies.
\item From their ground based observations Roche\etal (1991, see
also Moorwood 1986) found that many active galactic nuclei
(AGNs) show distinctively different
MIR spectra from those of starburst galaxies, in particular 
the broad band emission features are absent.
This is confirmed by new ISO observations (e.g. Lutz\etal 1997a;
Genzel\etal 1998).
Roche\etal (1991) argued that it is the AGN itself, instead of
other reasons such as the abundance of certain elements,
that affects the dust emission process to suppress the MIR features
either by alternating the grain composition within the nuclear
region, perhaps by the selective destruction of grain species, or
through very different excitation conditions.
\end{enumerate}

\subsection{Our model}
Our model is based on the following assumptions:

\nid{\it Assumption 1}:
The MIR SED (3--16\micron) of galaxies has three components, each
has a fixed SED given by the corresponding template (Figures~1--3):
(1) a cirrus/PDR (Photo-Dissociation Region)
component which is dominant
in the MIR SEDs of most of normal galaxies, (2) a starburst component due
to dust heated by the very intense radiation in starburst regions which is
likely to be the dominant component for violent starburst galaxies
such as Arp~220, and (3) an AGN component due to the dust emission
associated with AGNs. 

The adopted MIR spectrum (3--16\micron) of the cirrus/PDR component
is presented in Figure~1 (the solid curve).  Lu\etal (1998) gives two 
average spectra, one from a set of 10 ``FIR-cold'' galaxies with an
IRAS 60\micron--to--100\micron
flux density ratio, $f_{60\mu}/f_{100\mu}$, between 0.28 
and 0.40; and one from a set of 9 ``FIR-warm'' galaxies with $0.60 < 
f_{60\mu}/f_{100\mu} < 0.88$.   The relative flux accuracy of these spectra 
should be good to about 20\%.   The shape of our cirrus/PDR template 
spectrum at wavelengths shorter than $\sim$12\micron is identical to that
of the average spectrum for the FIR-cold galaxies in Lu\etal (1998).
At longer wavelengths a tail is taken from the spectrum of NGC~5195 
(Boulade\etal 1996). It can be seen that the two spectra agree 
reasonably well at the wavelengths covered by both.

In the wavelength range of 8.7--16\micron the spectrum of the starburst
component (Figure~2) is taken from the ISO-CAM CVF spectrum of the 
overlapping region (region A) in Arp~244 (Vigroux\etal 1996), 
the famous Antennae. This region is where the two disks crash, and 
where much of the star formation in this merging system is taking 
place.  The emission features are still visible in the spectrum, but
most of the energy is in the continuum which has a relatively steep 
slope. Shorter than 8.7\micron, the shape of the spectrum of region A in
Arp~244 is very close  to that of the average spectrum
for the FIR-warm galaxies in Lu\etal (1998). Therefore shorter than
8.7\micron the starburst component is taken from the latter (Fig.2) which
extends to 3\micron. Interestingly, longer than 8.7\micron 
the MIR continuum in the average spectrum of the FIR-warm galaxies
of Lu\etal is not as strong as in the starburst region in Arp~244,
suggesting that in these galaxies a significant fraction of the MIR
emission is emitted outside the starburst regions (i.e. associated
with the cirrus/PDR component). 

It should be noted that in the short wavelength 
channel of ISOPHT-S ($<$5\micron) the mean spectra of Lu\etal (1998)
are rather noisy, and show no significant evidence for any detection of 
features (e.g. the 3.3\micron feature)
at $\lambda <$5\micron. Thus both cirrus/PDR template and the
starburst template presented above are approximated by flat spectra
in the wavelength range 
3\micron$ < \lambda <$5\micron, determined by the average flux densities
(over the same wavelength range)  of the
mean spectrum of the FIR-cold galaxies and 
of the mean spectrum of the FIR-warm galaxies of Lu\etal (1998), 
respectively.

The template of the MIR SED of the AGN component (Figure~3) is based on the MIR
spectrum of the Seyfert 2 nucleus of NGC~1068 (Lutz\etal 1997a).
Because we are only interested in the broad band SED, the narrow
emission lines in this wavelength range are not included.

\nid{\it Assumption 2}: 
For a given galaxy with measured IRAS fluxes of 
$f_{25\mu}$ and $f_{60\mu}$, the 25\micron flux of the AGN component can be
determined using the following `color-method':
\begin{equation}
\rm f_{25\mu}^{AGN} = \cases{ f_{25\mu}
\;\; & $\rm (f_{60\mu}/f_{25\mu} \leq 2.5)$; \cr
f_{25\mu}\times {5 - f_{60\mu}/f_{25\mu}\over 5 - 2.5}
\;\; & $\rm (2.5 < f_{60\mu}/f_{25\mu} < 5)$; \cr
0 \;\; & $\rm (f_{60\mu}/f_{25\mu} \geq 5)$. \cr} \label{eq:agn1} 
\end{equation}

This assumption is based on the fact that galaxies with low 
$f_{60\mu}/f_{25\mu}$ ratios have a large chance of exhibiting Seyfert
activity (de Grijp\etal 1985). In an optical spectroscopic survey 
for a FIR color selected sample of 358 extragalactic
sources with $1 < f_{60\mu}/f_{25\mu} <3.72$
(i.e. $-1.5 < \alpha_{25,60} < 0$),  Keel\etal (1994) found that 
more than 60 percent of the sources are Active galaxies (including
Seyferts, 
QSOs and BL Lacs). In Figure~4 is plotted the $f_{25\mu}/f_{12\mu}$
vs. $f_{25\mu}/f_{60\mu}$ color-color diagram of the IRAS 25\micron
selected sample of Shupe\etal (1998a). Sources with known AGNs
(identified using NED) are plotted with different symbols.
Indeed we find that for
$\log (f_{25\mu}/f_{60\mu})\geq -0.4$ (i.e. $f_{60\mu}/f_{25\mu}\leq 2.5$)
the vast majority of the sources are Active galaxies. In particular,
most of the QSOs
are on the right side of this boundary. On the other hand, most of the normal
galaxies (i.e. without AGN) have $\log (f_{25\mu}/f_{60\mu})\leq -0.3$ (i.e. 
$f_{60\mu}/f_{25\mu}\geq 5$). Interestingly, many galaxies with AGNs,
in particular most of the LINERs, also have 
$\log (f_{25\mu}/f_{60\mu})\leq -0.3$. It is likely that for these
sources the IR fluxes are mainly from the host galaxies and the
AGN component is not significant.

Once $f_{25\mu}^{AGN}$ is determined, the 12\micron flux of the AGN 
component is derived as follows:

\begin{equation}
\rm f_{12\mu}^{AGN} = \cases{0.6\; f_{25\mu}^{AGN} 
\;\; & $\rm  (f_{25\mu}^{AGN} < f_{25\mu}$ and 
$\rm 0.6\; f_{25\mu}^{AGN} < f_{12\mu})$ \cr
f_{12\mu} 
\;\; & (otherwise) \cr } \label{eq:agn2} 
\end{equation}
The standard $f_{12\mu}/f_{25\mu}$ ratio of 0.6 for the AGN component
is taken from that of the Seyfert 2 nucleus of NGC~1068 (Lutz 1997a).

Based on the IRAS data, Giuricin\etal (1995) found that 
Seyfert 2 galaxies tend to have steeper continuum spectra between
12 and 25\micron than those of Seyfert 1 galaxies. However, given
the large beams of IRAS detectors, it is not clear whether this
reflects the intrinsic difference between the two types of AGN
or whether it is due to the difference in the star formation activity
around Seyfert 1s and Seyfert 2s (Maiolino\etal 1995).
In this work, we choose to use a single template (Figure~3) to model
the MIR SED of both types of AGN.

\nid{\it Assumption 3}: 
Define
\begin{equation}
\rm f_{25\mu}^{G} =  f_{25\mu} - f_{25\mu}^{AGN} 
\end{equation}
and
\begin{equation}
\rm f_{12\mu}^{G} =  f_{12\mu} - f_{12\mu}^{AGN}\;\; . 
\end{equation}
The 12\micron fluxes of the starburst component and of the cirrus/PDR
component can be determined by the color  
$f_{25\mu}^{G}/ f_{12\mu}^{G}$ through the following formulae:
\begin{equation}
\rm f_{12\mu}^{SB} = \cases{f_{12\mu}^{G}
 \;\; & ($f_{25\mu}^{G}/ f_{12\mu}^{G}\geq 6$); \cr
 & \cr
{f_{25\mu}^{G}/ f_{12\mu}^{G} - 2\over
6 - 2} \;\; 
& ($\rm 2 < f_{25\mu}^{G}/ f_{12\mu}^{G} < 6$); \hfil \cr
 & \cr
0 \;\; & ($f_{25\mu}^{G}/ f_{12\mu}^{G} \leq 2 $) \cr} \label{eq:sb} 
\end{equation}
and
\begin{equation}
 \rm f_{12\mu}^{PDR} = f_{12\mu}^{G}-f_{12\mu}^{SB}. \label{eq:pdr} 
\end{equation}
\oneskip

In this assumption, we assign a significantly higher standard
$f_{25\mu}/ f_{12\mu}$ color ratio to the starburst component 
($f_{25\mu}^{SB}/ f_{12\mu}^{SB}=6$) than that
of the cirrus/PDR component ($f_{25\mu}^{PDR}/ f_{12\mu}^{PDR}=2$), 
to be consistent with the fast-rising MIR SED of the starburst
template (Figure~2). This is also consistent with the anti-correlation
between $f_{12\mu}/ f_{25\mu}$ and $f_{60\mu}/ f_{100\mu}$ of galaxies
(Helou 1986) given that the $f_{60\mu}/ f_{100\mu}$ ratio is a good star
formation strength indicator (Bothun\etal 1989; Xu and De Zotti 1989).
For example Arp~299 (NGC~3690/IC~694), 
a famous local example of galaxy mergers,
has $f_{25\mu}/ f_{12\mu}=5.8$ and $f_{60\mu}/ f_{100\mu}=0.95$.

Given Equations \ref{eq:agn1}, \ref{eq:agn2}, \ref{eq:sb}, and
\ref{eq:pdr}, we can calculate the 12\micron fluxes of the
three components for a given galaxy with measured 
$f_{12\mu}$, $f_{25\mu}$ and $f_{60\mu}$. These 12\micron
fluxes provide the absolute
calibrations for the MIR SED templates of these components,
the sum of which is the model MIR SED (3--16\micron)
for the galaxy in question. 

\nid{\it Assumption 4}
There are no broad band features at wavelengths longer than 16\micron
in the IR emission of galaxies.

This assumption is based on the available IR spectra in the literature.
In particular the ISOSWS spectrum of M~82 (Lutz\etal 1997b), which
covers the wavelength range 3--40\micron, does not show
any broad band features beyond the 12.7\micron feature.
Consequently, for a given galaxy, the IR radiation spectrum  
at wavelengths longer than 16\micron is determined in our SED model by 
fitting the 3 IRAS fluxes at 25, 60 and 100\micron (taking into account
the real responsivity functions of IRAS detectors), and
the flux density at 16\micron predicted by the sum of the three MIR components,
using a simple cubic spline algorithm
(obviously the continuum so determined joins at 16\micron the model MIR SED).
The advantage of this empirical model is that it preserves the observed
flux measurements of each source and allows the inclusion of
a large sample of galaxies in a wide range of observed SEDs.

Figure~5 is a flow chart showing how actually our model proceeds.

\subsection{Tests of the SED model}
In Figure~6 the predicted MIR spectrum of M~82 and the observed ISOSWS
spectrum of M~82 (Lutz\etal 1997b) are compared. Allowing for
the small apertures of ISOSWS channels (ranging from $14''\times 20''$
to $17''\times 40''$), we have multiplied the IRAS fluxes of M~82
at 12, 25, 60, and 100\micron by 0.38, 0.5, 0.5
and 0.38, respectively. Admittedly these factors are somewhat
arbitrary because at these wavelengths no high resolution maps of M~82
are available for us to do aperture photometry. So they are chosen
to give the best fit. Nevertheless they look reasonable. In particular,
the two different factors (0.5 for $f_{25\mu}$and $f_{60\mu}$, 0.38 for
$f_{12\mu}$ and $f_{100\mu}$) can be understood by means of the known
anti-correlation in the $f_{12\mu}/f_{25\mu}$ vs. $f_{60\mu}/f_{100\mu}$
diagram (Helou 1986), which suggests that $f_{25\mu}$and $f_{60\mu}$
are more concentrated in the starburst nucleus of M~82.
The good agreement between the predicted and observed MIR {\it
continuum} is indeed due to the fine tuning of the aperture corrections,
while the good agreements between the shapes of 
the MIR {\it features} (which are independent of the fitting of
the continuum) on the predicted and observed spectra are truly remarkable.

We made a further test to determine whether the model 
reliably predicts galaxies' MIR SEDs.
Ashby\etal (1999) observed with ISOPHT-P a sample of 29 FIR bright
galaxies (f$_{60\mu}> 5.8$ Jy) in
several MIR wavelength bands, including
the PHT-P 16\micron band ($52''$ aperture).  
Our SED model was applied to a subsample of 18 
of the Ashby\etal galaxies.  The subsample was selected
based on four criteria: 
(1) the signal-to-noise ratio of the PHT-P 16\micron flux $>3$;
(2) detection of the galaxy by IRAS in both 
the 12 and 25\micron bands;
(3) optical extent $<2'$, or when in a galaxy pair the
combined size of the pair is $<2'$, 
(4) the distance to 
the nearest neighbor (of similar optical brightness) $>4'$. 
The last two criteria ensure that 
the PHT-P observations include most of the emission
from the region enclosed in the IRAS apertures ($\sim 4'$).
Convolving the predicted spectra with the PHT-P 16\micron 
responsivity function, the 16\micron fluxes are predicted 
and compared with the observed $f_{16\mu}$ (Figure~7). 
Given the uncertainty of the observed $f_{16\mu}$
(in most case $\sim 30\%$, Ashby et al. 1999), the agreement 
between the predicted and observed $f_{16\mu}$'s is 
quite good.
The model spectra of the same galaxies are plotted in Figure~8.

\subsection{The Sample}
The IRAS 25\micron selected sample (complete down to
$f_{25\mu}=0.25$~Jy) of Shupe\etal (1998a) contains
1456 galaxies, all have measured redshifts and measured 
IRAS 25\micron fluxes. In order to have a local sample
free from significant evolutionary effects,
our sample includes only the 1406 galaxies in
Shupe et al's sample which have redshifts $\leq 0.1$. 
Most of them are also detected by IRAS at 12, 60 and 100\micron.
At 12\micron, 330 (out of 1406) galaxies are
not detected by IRAS and have therefore  
only upper limits. For $f_{60\mu}$ and
$f_{100\mu}$, 18 and 94 galaxies have upper limits, respectively.
We assign a flux density $f_\nu =0.8\times$upper limit for these cases.

The SED model thus produces for each galaxy in the sample a SED from
3--120\micron. As examples, the SEDs of 50 galaxies with $9 \leq
\log (\nu L_\nu (25\mu{\rm m})/L_\sun) \leq 10$ are plotted in Figure~9.

\section{Local luminosity function at 15\micron}

Several MIR surveys (Oliver\etal 1997; Franceschini\etal 1996)
are being carried out by ISOCAM in the 15\micron band, 
the most sensitive one for starburst galaxies among ISOCAM filters.
However, there is no 15\micron local luminosity function 
(LLF) available in the literature, because there have been no 
complete samples of galaxies in 15\micron band.
Oliver\etal (1997) used the 12\micron LLF of Rush
et al. (1993) in the interpretation of their 15\micron deep survey.
This may be problematic because different galaxies have different 
15-to-12\micron flux ratios. 

We construct a 15\micron LLF in this section 
exploiting the model SEDs of the 1406 galaxies in our sample. 
First, for each galaxy in the sample, an in-band
flux at 15\micron 
is derived by convolving its SED with the responsivity functions of 
ISOCAM 15\micron filter (13--18\micron, ISOCAM User's Manual).
The monochromatic flux at 15\micron is derived from this by assuming
$\nu f_{\nu}=$constant within the bandpass.
When there is only an upper limit at 12\micron, 
instead of multiplying a by factor of 0.8 to get an approximation of
the real flux as in the previous section, here we input the 12\micron
upper limit into the SED model and obtain an upper limit for the 15\micron
flux. The 15\micron LLF can then be calculated
from the predicted 15\micron luminosities, 
and the corresponding upper limits, of these galaxies.

Since our sample is 25\micron-flux limited, not 15\micron-flux
limited, the bivariate method is used in the calculation of the 15\micron
luminosity function. The basic formula is
\begin{equation}
 \Psi(L_{15\mu} > l_{15\mu}) =
\int \Theta(L_{15\mu} > l_{15\mu} |L_{25\mu})\; \rho(L_{25\mu})\;\;
 d\, L_{25\mu}
\end{equation}
here $\Psi$ is the 15\micron {\it integral} luminosity function,
$\rho$ the 25\micron {\it differential} luminosity function
(Shupe\etal 1998a), and $\Theta$
the conditional probability function for a galaxy with given 25\micron
luminosity $L_{25\mu}$ to have the 15$\mu$ luminosity $L_{15\mu}$
brighter than $l_{15\mu}$. 
The integration is over the entire range of the 25\micron luminosity.
The derivative of $\Psi$ gives the differential luminosity function:
\begin{equation}
 \phi(L_{15\mu}) =
\int {\partial\Theta\over \partial L_{15\mu}} \rho(L_{25\mu})\;\;
 d\, L_{25\mu}\;\; .  \label{eq:lf}
\end{equation}

In practice, the luminosity functions are calculated in the domain of
the logarithms of the luminosities. Hence the $L_{15\mu}$'s and
the $L_{25\mu}$'s in above formulae
should be replaced by the corresponding logarithms.
The sample is binned for both $\log (L_{25\mu})$ and
$\log (L_{15\mu})$, and Eq(\ref{eq:lf}) can be approximated by
\begin{equation}
 \phi[\log (L_{15\mu})=L_i] =\sum_j P_{i,j} \rho[\log(L_{25\mu})=L_j]\;\; 
\Delta_j/\delta_i 
\end{equation}
where $L_j$ and $\Delta_j$ are the center and the width of the j-th
bin of $\log (L_{25\mu})$,
$L_i$ and $\delta_i$ are the center and the width of the i-th bin of 
$\log (L_{15\mu})$, and $P_{i,j}$ is the probability of  
galaxies in the j-th bin of $\log (L_{25\mu})$ 
(i.e. $L_j-0.5\Delta_j \leq log(L_{25\mu}) < L_j+0.5\Delta_j$)
to have $\log (L_{15\mu})$ in the bin $L_i$
($L_i-0.5\delta_i \leq \log (L_{15\mu})  < L_i+0.5\delta_i$).

In order to take into account the information contained in
the upper limits, the Kaplan-Meier estimator (Kaplan and Meier 1958)
is applied in the calculation of $P_{i,j}$  
and its variance. In what follows we give a brief description of the algorithm.

For the subsample of galaxies in the j-th bin of $\log (L_{25\mu})$, 
define $F_{i-1}$ as the probability function
\begin{equation}
F_{i-1}=P[\log(L_{15\mu})<L_i+0.5\delta_i]
\end{equation}
and 
\begin{equation}
F_{i}=P[\log(L_{15\mu})<L_i-0.5\delta_i]\;\; .
\end{equation}
So, 
\begin{equation}
P_{i,j}=F_{i-1}-F_{i}
\end{equation}
and
\begin{equation}
Var(P_{i,j})=Var(F_{i-1})+Var(F_i)-2\times Covar(F_{i-1},F_i)\;\; .
\end{equation}
The algorithm for calculating the KM estimator of
$F_{i-1}, F_i, Var(F_{i-1})$ and
$Var(F_i)$ can be found in Feigelson and Nelson (1985), while the
KM estimator for the covariance between $F_{i-1}, F_i$ (for 
$F_i > 0$) is (Akritas and LaValley 1997)
\begin{equation}
Covar(F_{i-1},F_i) = {F_{i-1}\over F_i} Var(F_i)\;\; .
\end{equation}
 
Finally, the variance of $\phi$ is
\begin{eqnarray}
 Var[\phi(\log(L_{15\mu})=L_i)] = \sum_j 
\{ & Var(P_{i,j})\rho[\log(L_{25\mu})=L_j] + \nonumber \\ 
& P_{i,j}Var[\rho(\log(L_{25\mu})=L_j)]  \}
(\Delta_j/\delta_i)^2 \; .
\end{eqnarray}

The resultant 15\micron differential luminosity function 
is given in Table 1 and plotted in Figure~10.
A least-squares fit to the {\it integral}
15\micron luminosity function
by a smooth function with the form (Yahil\etal 1991)
\begin{equation}
 \Psi(L) = C \left(L\over L_\star \right)^{-\alpha}
\left(1+L\over L_\star\right)^{-\beta}
\end{equation}
is carried out. The parameters of this fitting function are given in Table 2.
The {\it differential} 15\micron luminosity function specified by
these parameters is given in Figure~10 by the solid line.

\section{MIR number counts model and effects of emission features}

\subsection{Model for number counts}
Our model for galaxy counts as a function of the
limiting MIR flux and redshift follows the prescription of Condon
(1984; see also Hacking\etal 1987; Hacking \& Soifer 
1991). Here we give a brief outline of the formulation. Details 
can be found in the above references. 

We consider only one set of cosmological parameters, namely
$H_0=75$, $\Omega = 1$ ($q=0.5$), and the cosmological constant
$\Lambda=0$. 

Let $\rho (L)$ be the local differential
luminosity function (number of galaxies in
unit comoving volume and unit $\log(L)$ interval).
When the evolutionary effects are taken into account, the LF
at redshift $z$ is 
\begin{equation}
\rho' (L,z) = G(z)\, \rho \left( {L\over F(z)} \right);
\end{equation}
where $G(z)$ is the density evolution function and $F(z)$ the luminosity
evolution function. 

The monochromatic luminosity in the rest frame is: 
\begin{equation}
 L_\nu = f_\nu 4\pi D_{L}^2 \times 10^{0.4K}, \label{eq:l}
\end{equation}
where $f_\nu$ (hereafter f) is
the observed flux density, $z$ the redshift, 
$D_L$ the luminosity distance:
\begin{equation}
D_L={2c\over H_0}\; (1+z)[1-(1+z)^{-0.5}]\;\; , 
\end{equation}
and $K$ the standard
K-correction (Lang 1980):
\begin{equation}
K=2.5\log(1+z)+2.5\log{\int_{\lambda_1}^{\lambda_2} 
S(\lambda)R(\lambda)\; d\lambda\over \int_{\lambda_1}^{\lambda_2} 
S(\lambda/(1+z))R(\lambda)\; d\lambda} \; .  \label{eq:k}
\end{equation} 
The derivative of the comoving volume vs. redshift $z$ is
\begin{eqnarray}
{dV\over dz} & = & 4\pi D_A^2 \;\; {d\; D_A\over d\; z} \nonumber \\
       & = & 4\pi D_A^2 \; {c\over H_0(1+z)^{1.5}} \; . \label{eq:dv}
\end{eqnarray} 
where $D_A$ is the angular distance:
\begin{equation} 
D_A= {2c \, [1-(1+z)^{-0.5}] \over H_0} \;\; .
\end{equation} 

By definition the K-correction is a function of the redshift. 
However, for a given bandpass ($R(\lambda)$), the K correction 
also depends on the detailed spectrum ($S(\lambda)$),
which varies from galaxy to galaxy. Therefore for the model
it is more
appropriate to define the K-correction as a random variable with its
{\it probability distribution} depending on the redshift and on some other 
observables. Our prescription for the K-correction explicitly includes
a luminosity dependence because IRAS data have shown
that the $f_{12\mu}/f_{25\mu}$ color (and therefore the MIR SED)
is a strong function of the MIR luminosity (Fang\etal 1998).
In this work we choose to approximate the K distribution by a Gaussian:
\begin{equation}
U(K,L,z) = {1\over \sqrt{2\pi}\sigma_K} \;
\exp{-(K-K_0)^2\over 2\sigma^2_K} \label{eq:u}
\end{equation}
where the mean $K_0$ and the dispersion $\sigma_K$ 
are dependent on the {\it luminosity} $L$ and the redshift $z$.
We have also tried a prescription where 
the K-correction is a function of the 
FIR-to-blue luminosity ratio instead of the MIR luminosity
$L$. No significant changes are found in the results for the number counts.

Given the above definitions, we have:
\begin{equation}
\eta (L,z,K)\; d\log (L)\; dz\; dK  =  \rho' (L,z)
\; U(K,L,z)\; {dV\over dz}\; d\log(L)\; dz \; dK
\end{equation}
where $\eta$ is the counts per unit $\log(L)$ interval per unit 
z interval per unit $K$ interval.
Substituting the variable $L$ by $f$ according to Eq(\ref{eq:l})
for given $z$ and K, and integrating over K, one obtains the
prediction for the differential
counts per unit $\log(f)$ interval and per unit $z$ interval:
\begin{equation}
\xi (f,z)  =  \int_{-\infty}^{\infty} 
\rho' [(L(f,z,K),z]
\; U[K,L(f,z,K),z] {dV\over dz} \; dK \;\; . \label{eq:xi}
\end{equation}

\subsection{K-corrections}
The narrower the bandpass\footnote{The width of a bandpass
is better defined by $log(\lambda_2) - log(\lambda_1)$, which is not affected
by the redshift.}, the more significant the effect of the
emission features on the K-correction. Here we compare the predicted
K-corrections for three bandpasses: ISOCAM 12\micron band, 
ISOCAM 15\micron band (ISOCAM User's Manual), 
and WIRE 25\micron band (Shupe et al. 1998b). 
The responsivity of these filters are plotted in Figure~11. 

For each of the three bands, K-corrections at different $z$ are
calculated for each of the 1406 galaxies in our sample using its model
spectrum and Eq(\ref{eq:k}). Then the sample is binned according
to the monochromatic luminosity (rest frame) at the effective
wavelength of the band, and the means and the dispersions of the K-corrections
at different $z$ are calculated for the subsample of galaxies in
each bin, which are used as estimators of the $K_0(L,z)$ and
$\sigma_K (L,z)$ in Eq(\ref{eq:u}).

In Figures~12a, 12b and 12c the mean $K_0$ as function of
redshift $z$ for galaxies with different luminosities are plotted
for the three filters, respectively, and the dispersions $\sigma_K$ 
are plotted as error bars. And in Figures~12d, 12e and 12f,  
the dependences of $K_0$ and $\sigma_K$ on luminosity 
are plotted for $z=0.5$, $z=1$ and $z=2$ for the same filters.
The K-$z$ curves for WIRE 25\micron band flux 
demonstrate two dips near $z=1$ and $z=2$, which are due to the 
MIR emission features at $\sim 11$--12\micron and at 
$\sim 6$--8\micron (see Figures~1 and 2), 
respectively. The peak around $z=1.5$ is
caused by the trough at $\sim 9$--10\micron in MIR SEDs of
galaxies, which may partially be due to the silicate absorption. 
For the ISOCAM 15\micron band flux, the bump caused by this 
SED trough moves to $z\sim 0.5$ and the emission features 
at $7$--8\micron cause a dip around $z=1$. The K-$z$ curves for
the ISOCAM 12\micron band (the most broad band among the three)
are relatively smooth.  

\subsection{Number counts}
Through K-corrections, the MIR features affect the number counts and
the redshift distributions of deep surveys. 
Figures~13a, 13b and 13c show
Euclidean normalized differential counts predicted by different models
of the cosmological evolution of galaxies
for ISOCAM 12\micron, ISOCAM 15\micron and WIRE 25\micron bands. 
The $L\propto (1+z)^3$ model is chosen because it fits 
the IRAS 60\micron counts (Pearson and Rowan-Robinson 1996) well,
and the results from the $\rho\propto (1+z)^4$ model are presented to 
highlight the difference between a density evolution model and a luminosity 
evolution model. The latest local luminosity functions (LLFs) at 12\micron
(Fang\etal 1998) and 25\micron (Shupe\etal 1998a) which improve
upon the earlier LLFs in these bands (Soifer and Neugebauer 1991;
Rush\etal 1993), and the 15\micron LLF
presented in Section 3 are used in the calculations of model predictions.

In order to demonstrate the effects of the MIR features on the 
number counts, in Figures~13d, 13e and 13f we show the
Euclidean normalized differential counts predicted by evolution models
based on model spectra of galaxies {\bf without} the MIR features,
which are calculated using the smooth fits 
(power-laws) of the templates of the three components in Section 2.

In all of the three bands considered, the effects of the MIR features on
the number counts predicted by the density evolution model 
($\rho\propto (1+z)^4$) are significant. In Figures~13d, 13e, and 13f 
(no MIR features) the curves of predicted number counts 
by the density evolution model are rather flat, while in
Figures~13a, 13b, and 13c the corresponding curves show dips and bumps
caused by the MIR features.  For example in Figure~13b (15\micron)
the curve of number counts predicted by the density evolution model 
shows two shallow bumps which are due to the emission features
in the 11--13\micron wavelength range and those in the 6--8.5\micron 
wavelength range, respectively. The prominent bump at $f_{25\mu}
\sim 0.1$~mJy in Figure~13c (25\micron) is due to the features in 
both the 6--8.5\micron wavelength range 
and the 11--13\micron wavelength range (see the discussion of Figure~14).
It appears that in all three bands the MIR features cause only
slight distortions on the number counts predicted  
by the no evolution models.

There is a broad peak on the number counts predicted by 
the luminosity evolution model ($L\propto (1+z)^3$) 
in all six panels including those for models without the emission features.
Hence this peak is not due to the emission features. 
Instead it is a combined effect caused by other two factors,
one related to the evolutionary model and the other to the adopted 
cosmological model ($q_0=0.5$): The strong evolution ($L\propto (1+z)^3$) 
ensures that before the depth of a survey getting close to the `edge' of the 
visible universe (i.e. $z\lsim 1$) the number counts
increase much faster than the Euclidean predictions. Then, when
the flux level becomes so faint that the characteristic
redshifts of galaxies are significantly larger than unity
(for a given flux, the stronger the luminosity evolution, the larger
the characteristic redshift),
the slope of $dV/dz$ (Eq(\ref{eq:dv})) starts to decrease rapidly
with increasing $z$ (decreasing flux), 
and the number counts drop significantly below the Euclidean predictions.

The effects of the MIR features on the number counts predicted by
the luminosity evolution model look moderate. The peaks in Figure~13a
(12\micron) and in Figure~13b (15\micron) are slightly narrower than those 
in Figure~13d and Figure~13e, and the peak in Figure~13c (25\micron) is slightly
broader than that in Figure~13f.

In Figures~14a--f, we compare the redshift distributions
of sources at the flux level of $f_\nu=0.5$~mJy, predicted by
different evolutionary models. Note that these distributions are
for galaxies {\bf at} this flux level, rather than for galaxies in
samples brighter than this flux.  The flux $f_\nu=0.5$~mJy is
about the location of the peaks on the predicted number counts of
the luminosity evolution model (Figure~13), and it is close to the flux
limits of all ISOCAM deep surveys (Rowan-Robinson\etal 1997;
Franceschini\etal 1996) and that of the planned WIRE survey
(Hacking\etal 1996).

In all the three bands considered, the no evolution model
(both with and without MIR features) predicts the mode of redshift distribution
of galaxies with $f_\nu=0.5$~mJy at $z\sim 0.2$--0.4. This highlights
the reason why the effects of the MIR features are so weak
on the number counts predicted by this model:
because the high redshift galaxies never contribute significantly
to the number counts, the K-correction (through which the
MIR features affect number counts) is never significant.

On the other hand, the MIR features add prominent dips and bumps
on the redshift distributions predicted by the density evolution
model. For example the sharp peak at $z\sim 1$ on the dotted line
(density evolution model) in Fig.14c (25$\mu m$) is due to the features in 
in the 11 --- 13$\mu m$ wavelength range. This peak (after Euclidian 
normalization) reaches its maximum at the flux level when the 
11 --- 13$\mu m$ features
of the $L_\star$ galaxies are redshifted into the 25\micron filter
($L_\star$ does not change with redshift in the density evolution model,
hence for a given flux the $L_\star$ galaxies have the same redshift).
This happens at $f_{25\mu m}\sim 0.1 mJy$, and therefore
a strong bump appears at this flux level 
on the 25\micron number counts predicted by the density evolution model
(Fig.13c). At fainter flux level ($f_{25\mu m}\sim 30 \mu Jy$) the
bump is taken over by sources of $z\sim 2$ for which the strong
emission features in the 6--8.5\micron wavelength range are in
the bandpass of the WIRE 25$\mu m$ camera.

In Figures~14d, 14e and 14f (no MIR features), 
redshift distributions predicted by the luminosity evolution model
are much broader than those predicted by the density evolution
model and by the no evolution model. The reason for
this difference is the following:  When the K-correction is 
insignificant, the shape of the redshift distribution is
essentially determined by the product
of the luminosity function and the $dV/dz$ (c.f. Eq(\ref{eq:xi})).
Thus the redshift distributions predicted both by the density evolution
model and by the no evolution model
resemble the `visibility function' of a local
sample (e.g. Shupe\etal 1998a), with a sharp peak
centered at the redshift of the $L_\star$ galaxies (for the given flux).
On the other hand, in the luminosity evolution model, 
the $L_\star$ increases rapidly with redshift, causing a much broader
peak in the redshift distribution (e.g. Figure~14f). It is because of 
this broadness of the redshift distributions that the 
effects of the
MIR features on the number counts predicted by the luminosity
evolution model (Figure~13) are relatively weak, in particular compared
to those predicted by the density evolution model (Figure~13):
the effects are largely smeared out by including galaxies in
a wide range of redshifts for a given flux.

\section{Discussion}
MIR emission features, if they exist in the IR SEDs of
distant galaxies as in those of the local galaxies,
can cause  significant effects on the K-corrections (Figure~12),
especially when the bandpass 
is comparable to the widths of the features (e.g. ISOCAM 15\micron
and WIRE 25\micron bands). This
affects the predictions on MIR number counts and the
redshift distributions of different evolution models.
The dips-and-bumps caused by the MIR features on the
curves of the Euclidean normalized number counts  will provide
very useful diagnostic indicators of the type of 
evolution that is occurring, which
may help to distinguish the density evolution model from luminosity
evolution model when the redshifts are not 
available. On the other hand, the   
fluctuations in the slope of the number counts caused by the emission features
should be taken into account in any attempt of determining the
evolutionary rate from the {\it slope}. 

Dramatic features are caused by the MIR features on the redshift
distribution, suggesting that certain selection effects in redshift
space are expected in MIR surveys. This may indeed be beneficial
rather than annoying. For example, because of the strong emission
features at $\sim 6$--8\micron, the WIRE 25\micron survey may
detect a large number of galaxies of $z\lsim 3$.
However, in studies on large scale structures
based on deep MIR samples, these features must be separated from those
caused by large scale inhomogeneities.

\section{Summary and Conclusions}

Our main results are summarized as follows:

\begin{enumerate}
\item An empirical,
three-component model for the MIR SEDs of galaxies is developed.
The templates of (1) a cirrus/PDR component, (2) 
a starburst component and (3) an AGN component 
are constructed from published 
ISO observations of MIR spectra of galaxies. Tests show that the
model can reproduce the MIR spectrum of M~82 and the ISOPHT-P 16\micron
fluxes of a sample of galaxies quite well.
\item The model is applied to 
a sample of 1406 local galaxies ($z \leq 0.1$; Shupe et al. 1988a). 
Treating the K-correction as a random variable, its probability distribution
(approximated by Gaussian) as a 
function of the redshift $z$ and of the monochromatic
luminosity $L_\nu$ are calculated for three different bandpasses
(WIRE 25\micron, ISOCAM 15\micron and ISOCAM 12\micron)  
using these 1406 spectra. 
\item Effects of MIR emission features on the MIR source counts and on
the redshift distributions are studied for the same MIR bands assuming
three different evolution models (no-evolution, $L\propto (1+z)^3$
and $\rho\propto (1+z)^4$). It is found that predictions by 
luminosity evolution model ($L\propto (1+z)^3$) and 
density evolution model ($\rho\propto (1+z)^4$) for source counts and 
redshift distributions in the WIRE 25\micron band and
the ISOCAM 15\micron band are affected by the MIR emission
features significantly, showing dips and bumps which can be 
identified as being caused by specific MIR emission features. 
\item We point out that the dips-and-bumps on curves of MIR
number counts will be useful indicators  
of evolution mode. The strong emission
features at $\sim 6$--8\micron will help the detections
(especially in the WIRE 25\micron survey) of relatively high redshift
($z\sim 2$) galaxies. On the other hand determinations of the evolutionary
rate based on the slope of source counts, and studies on the large scale
structures using the redshift distribution of MIR sources will
have to treat the effects of the MIR emission features carefully.
\item A 15\micron local luminosity function is calculated from the
predicted 15\micron fluxes of the 1406 galaxies using the bivariate 
(15\micron vs. 25\micron luminosities) method. This luminosity
function will improve our
understanding of the ISOCAM 15\micron surveys.
\end{enumerate}
 

\vskip2cm 
We are grateful to James Peterson for providing 
the responsivity curve of WIRE 25\micron camera in electronic
form. CX thanks
Michael Akritas and Eric Feigelson for stimulating discussions
on the Kaplan Meier estimator, and Francois-Xavier D\'esert
for helpful comments on an earlier version of this paper.
NED is supported by NASA at IPAC. This work is supported in
part by a grant from NASA's Astrophysics Data Program.


\clearpage

\begin{deluxetable}{ccc}
\tablewidth{0pt}
\tablecaption{Local differential luminosity function at 15~$\mu$m
\label{tab:lf15}
}
\tablehead{\colhead{$\log(\nu L_{\nu} (15\mu m)/L_\sun )$} 
& \colhead{$\rm \log(\phi /(Mpc^{-3}mag^{-1}))$} & \colhead{1 $\sigma$ error}
}
\startdata
     7.7 &  -2.23  &   -2.82 \nl
     8.1 &  -2.88  &   -3.34 \nl
     8.5 &  -2.76  &   -3.43 \nl
     8.9 &  -2.88  &   -3.67 \nl
     9.3 &  -3.13  &   -4.06 \nl
     9.7 &  -3.52  &   -4.46 \nl
     10.1 &  -4.18  &   -5.05 \nl
     10.5 &  -5.06  &   -5.80 \nl
     10.9 &  -5.93  &   -6.58 \nl
     11.3 &  -7.00  &   -7.46 \nl
     11.7 &  -7.91  &   -7.91 \nl
\enddata      
\end{deluxetable}
\vskip2truecm

\begin{deluxetable}{cccc}
\tablewidth{0pt}
\tablecaption{Fitted Parameters of the 15~$\mu$m Luminosity Function 
\label{tab:para_lf15}
}
\tablehead{\colhead{$\alpha$} & \colhead{$\beta$} & 
\colhead{$\log(L_\star/L_\sun)$ } & \colhead{ $\rm \log (C/Mpc^{-3})$}
}
\startdata
$    0.47\pm 0.13 $ & $ 2.20\pm 0.13$ & $  9.62 \pm 0.15 $ &
    $-2.93773\pm 0.23 $ \nl
\enddata      
\end{deluxetable}


\clearpage

\figcaption[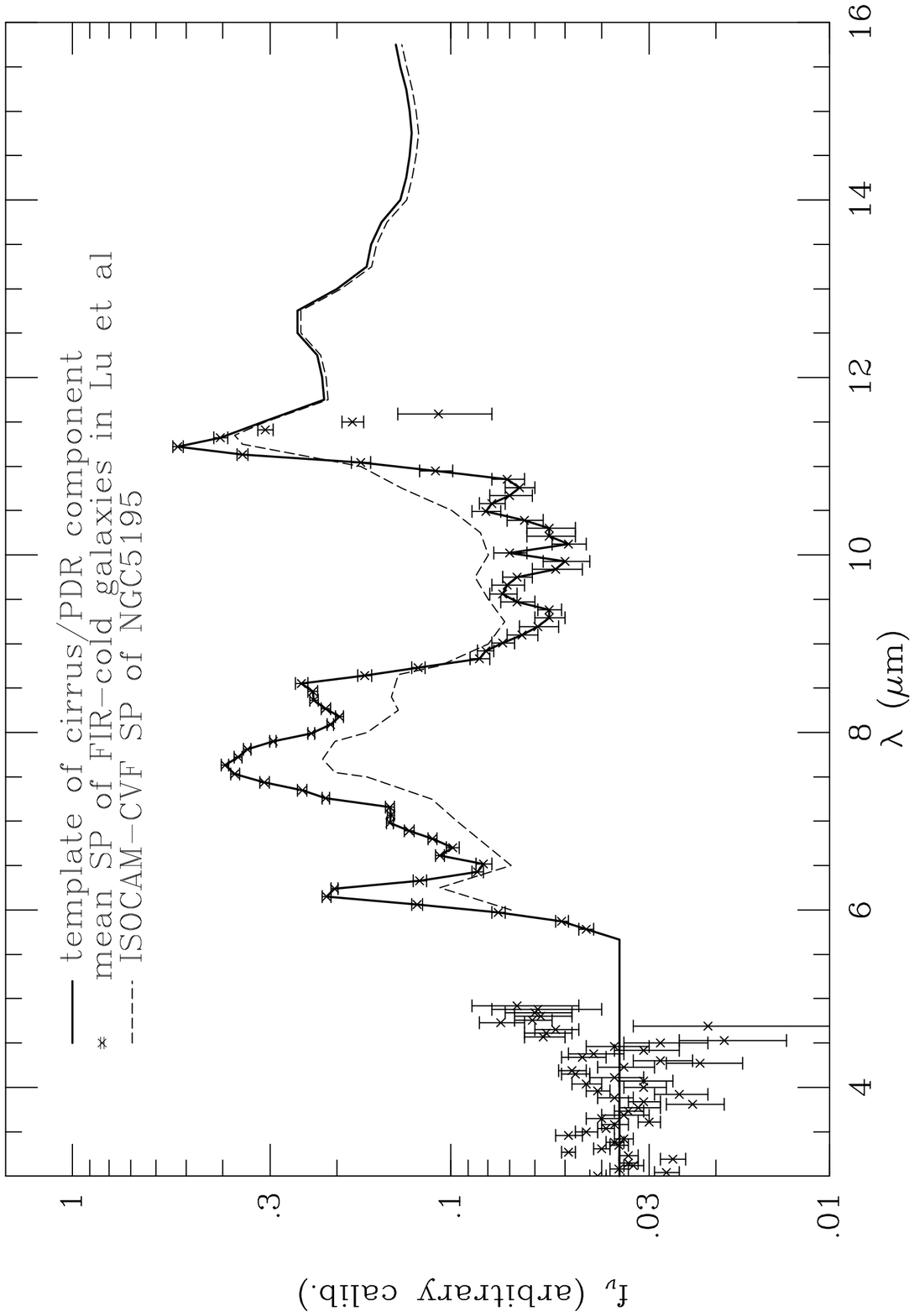]{
\underbar{{\it Solid curve}}: template of the MIR spectrum of
the {\bf cirrus/PDR} component.
\underbar{{\it Crosses with error bars}}: mean ISOPHT-S spectrum 
of the ten FIR cold galaxies in the sample of Lu{\it et al.} (1998).
\underbar{{\it Dashed curve}}: ISOCAM CVF spectrum of
NGC~5195 (Boulade{\it et al.} 1996).
\label{fig:pdrmir}
}

\figcaption[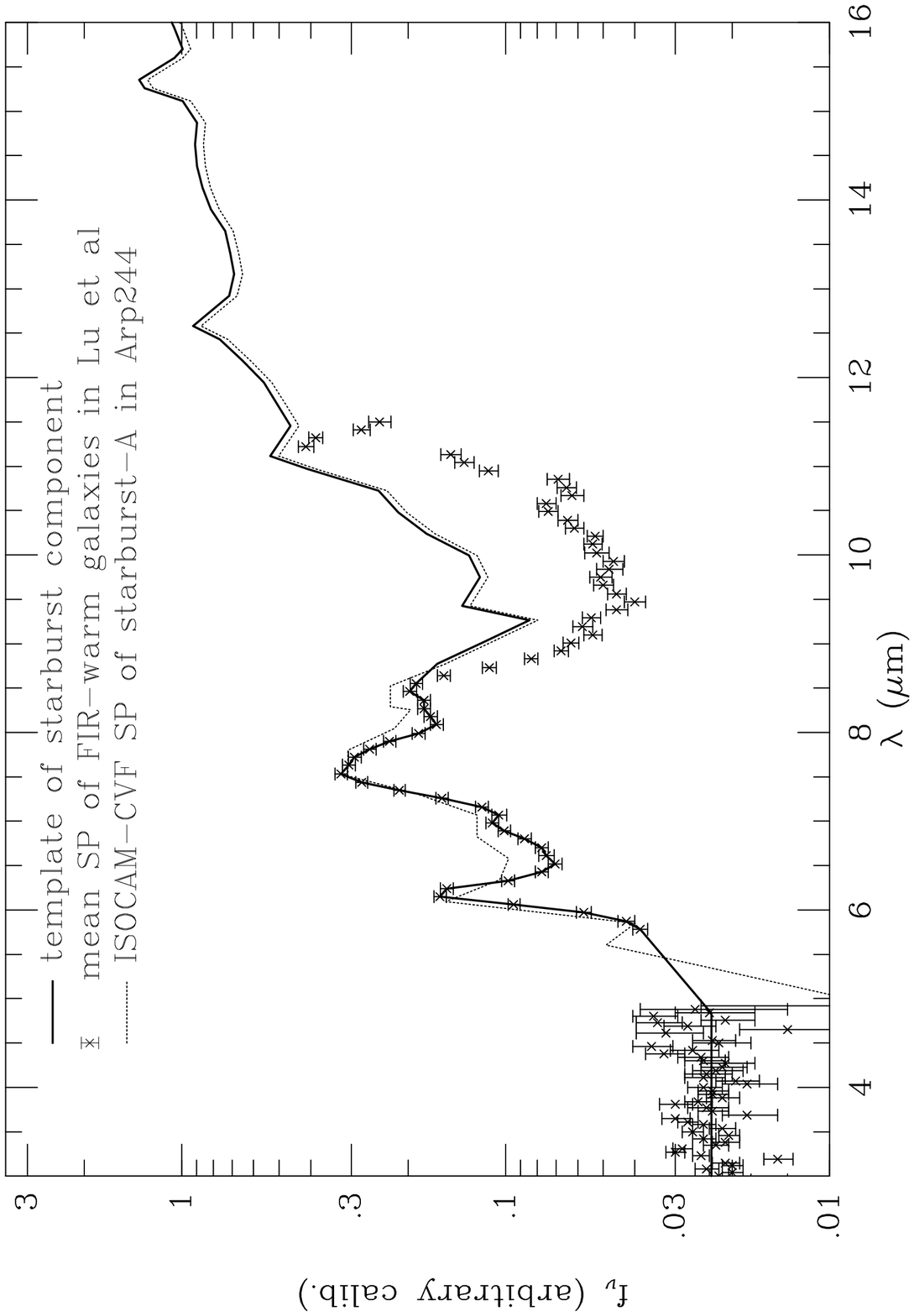]{
\underbar{{\it Solid curve}}: template of the MIR spectrum of
the {\bf starburst} component.
\underbar{{\it Crosses with error bars}}: mean ISOPHT-S spectrum of
of the nine FIR warm galaxies in the sample of Lu{\it et al.} (1997).
\underbar{{\it Dotted curve}}: ISOCAM CVF spectrum of
star-burst region A in Arp~244 (Vigroux{\it et al.} 1996).
\label{fig:sbmir}
}

\figcaption[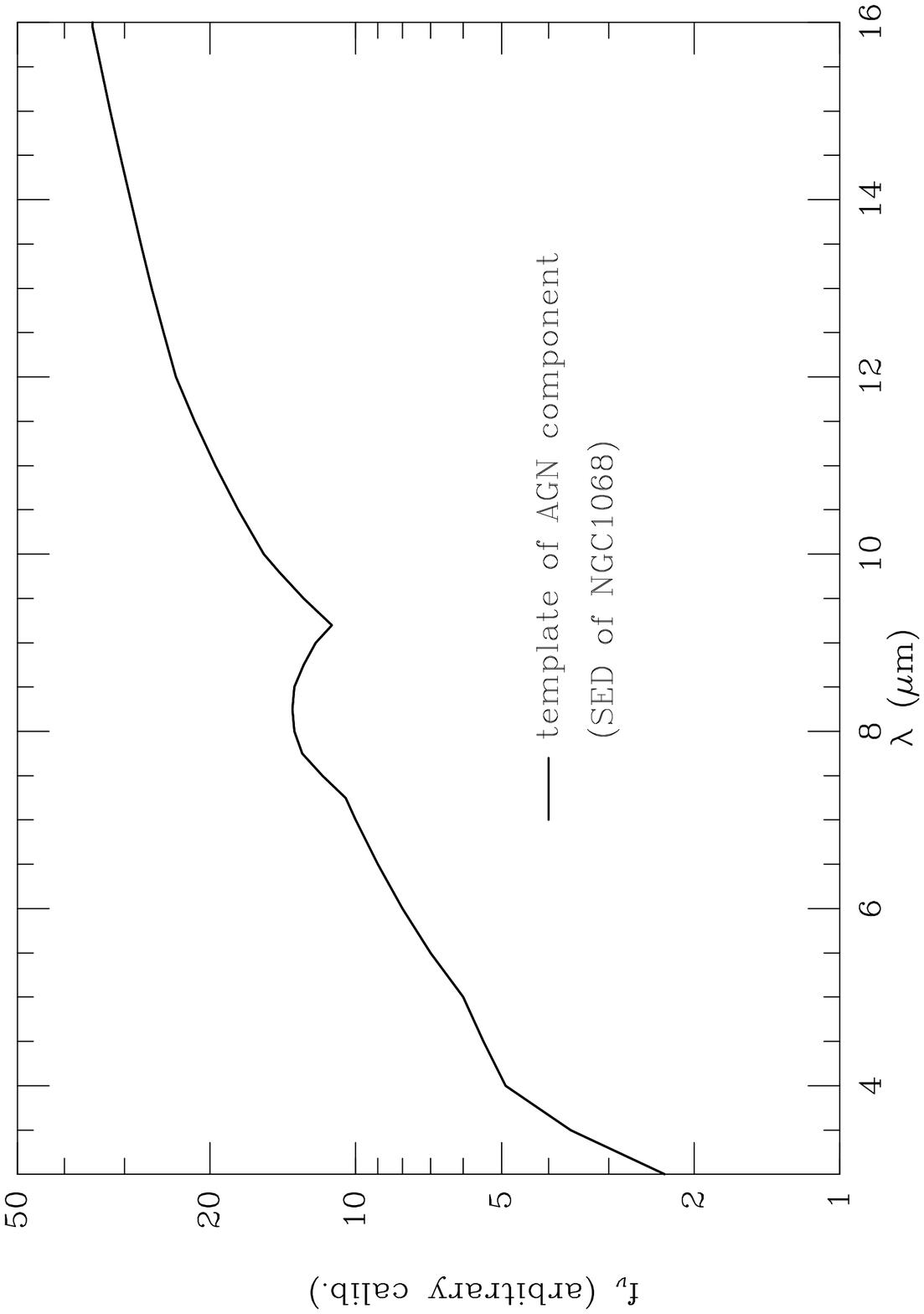]{
Template of the MIR spectrum of the {\bf AGN} component,
based on the MIR
spectrum of the Seyfert 2 nucleus of NGC~1068 (Lutz{\it et al.} 1997a).
Because we are only interested in the broad band SED, the narrow
emission lines are not included.
\label{fig:agnmir}
}

\figcaption[fig4.eps]{
The $f_{25\mu}/f_{12\mu}$
vs. $f_{25\mu}/f_{60\mu}$ color-color diagram of the IRAS 25~$\mu$m
selected sample of Shupe{\it et al.} (1998a). Sources with known AGNs
(identified through NED) are plotted with different symbols. The
rest of the sample are plotted by small dots. The dashed line
marks the adopted boundary beyond which all the IR emission is
considered to be due to the AGN component.
\label{fig:agncol}
}

\figcaption[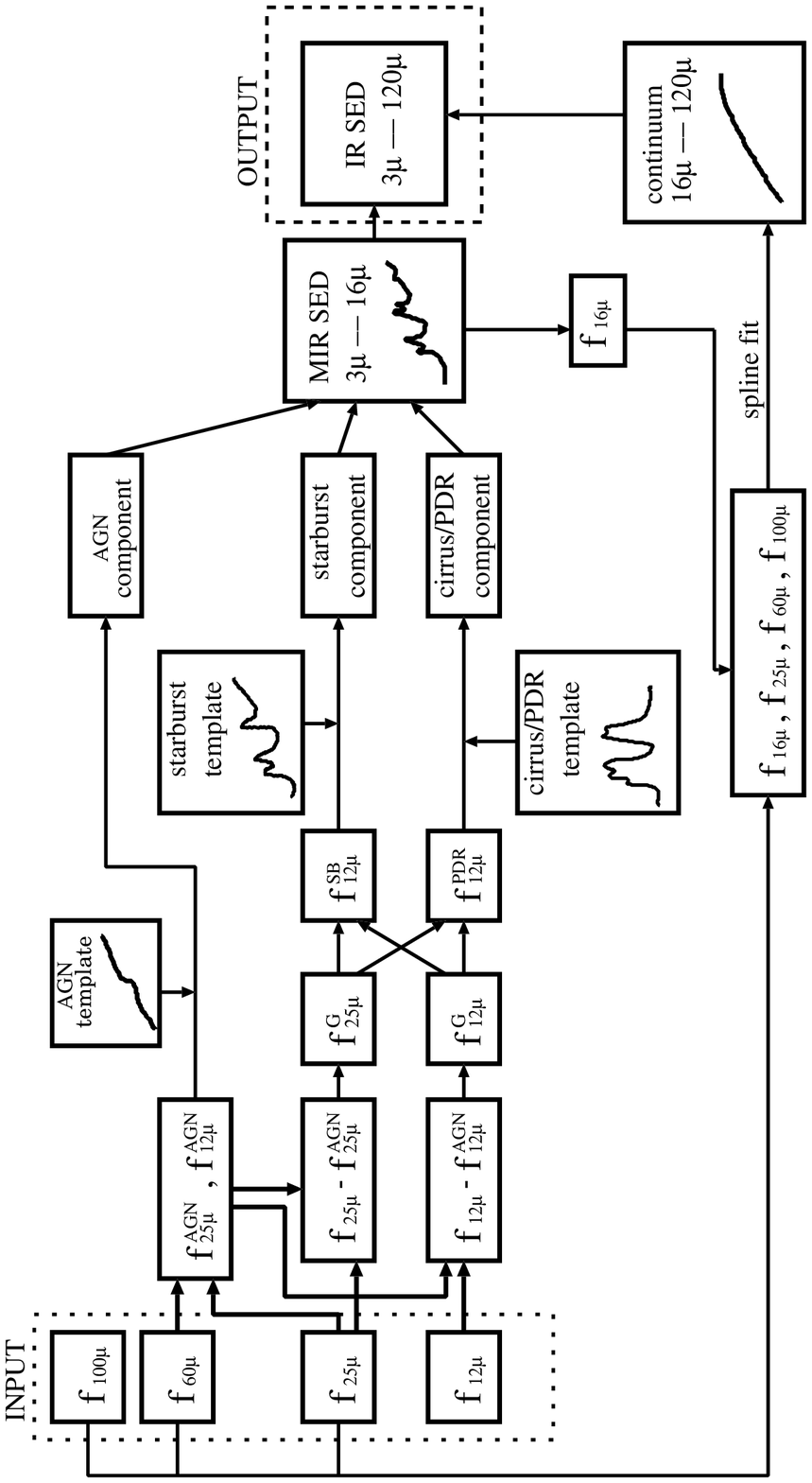]{
Running chart of the SED model.
\label{fig:sedmod}
}

\figcaption[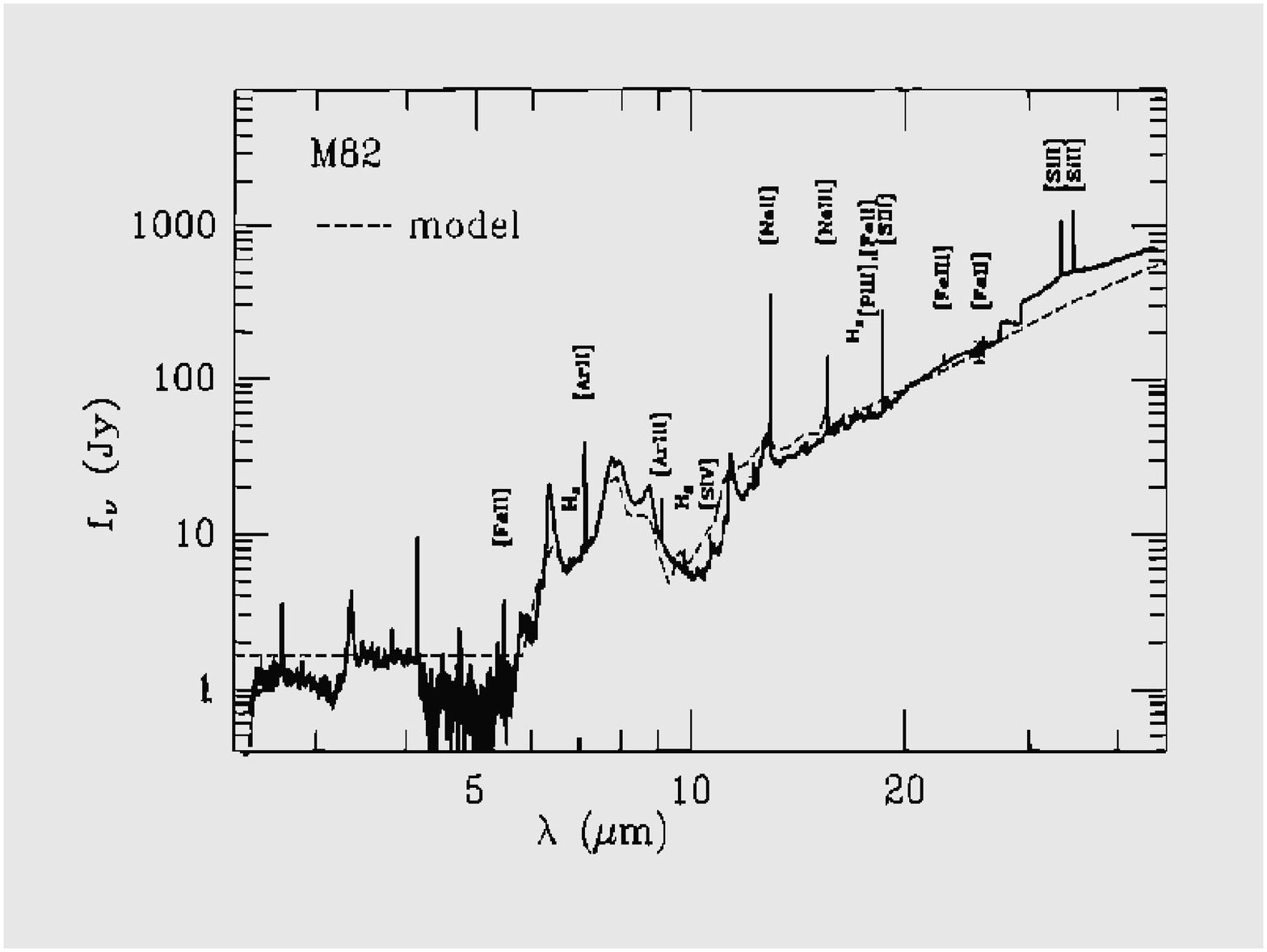]{
Comparison between predicted (dotted curve) and observed (solid curve,
taken from Lutz{\it et al.} 1997) MIR spectra of M~82.
\label{fig:m82}
}

\figcaption[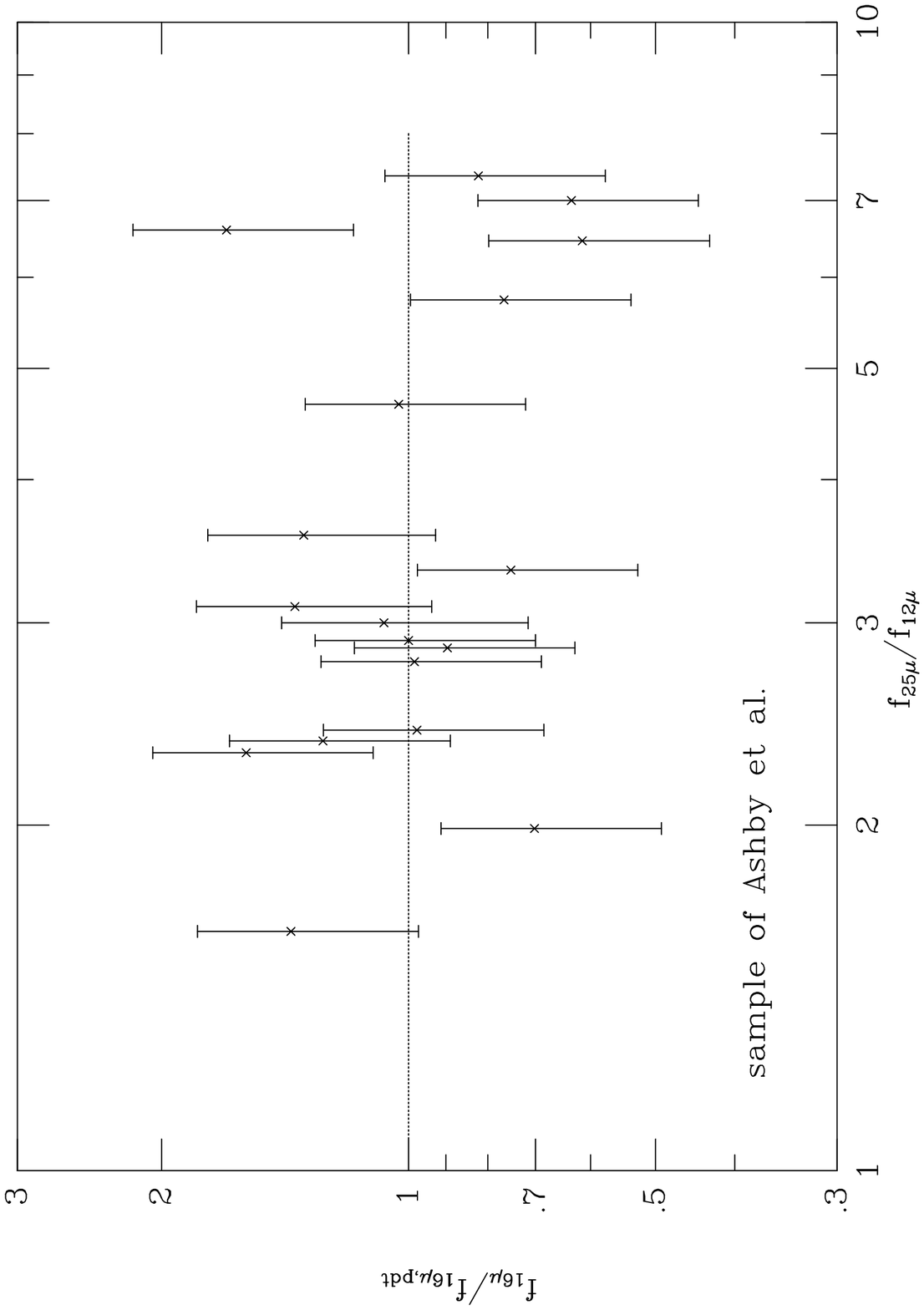]{
Comparisons between predicted and observed (ISOPHT) 16~$\mu$m fluxes
($f_{16\mu,prt}$ and $f_{16\mu}$, respectively) for
galaxies in the sample of Ashby{\it et al.}
\label{fig:f16comp}
}

\figcaption[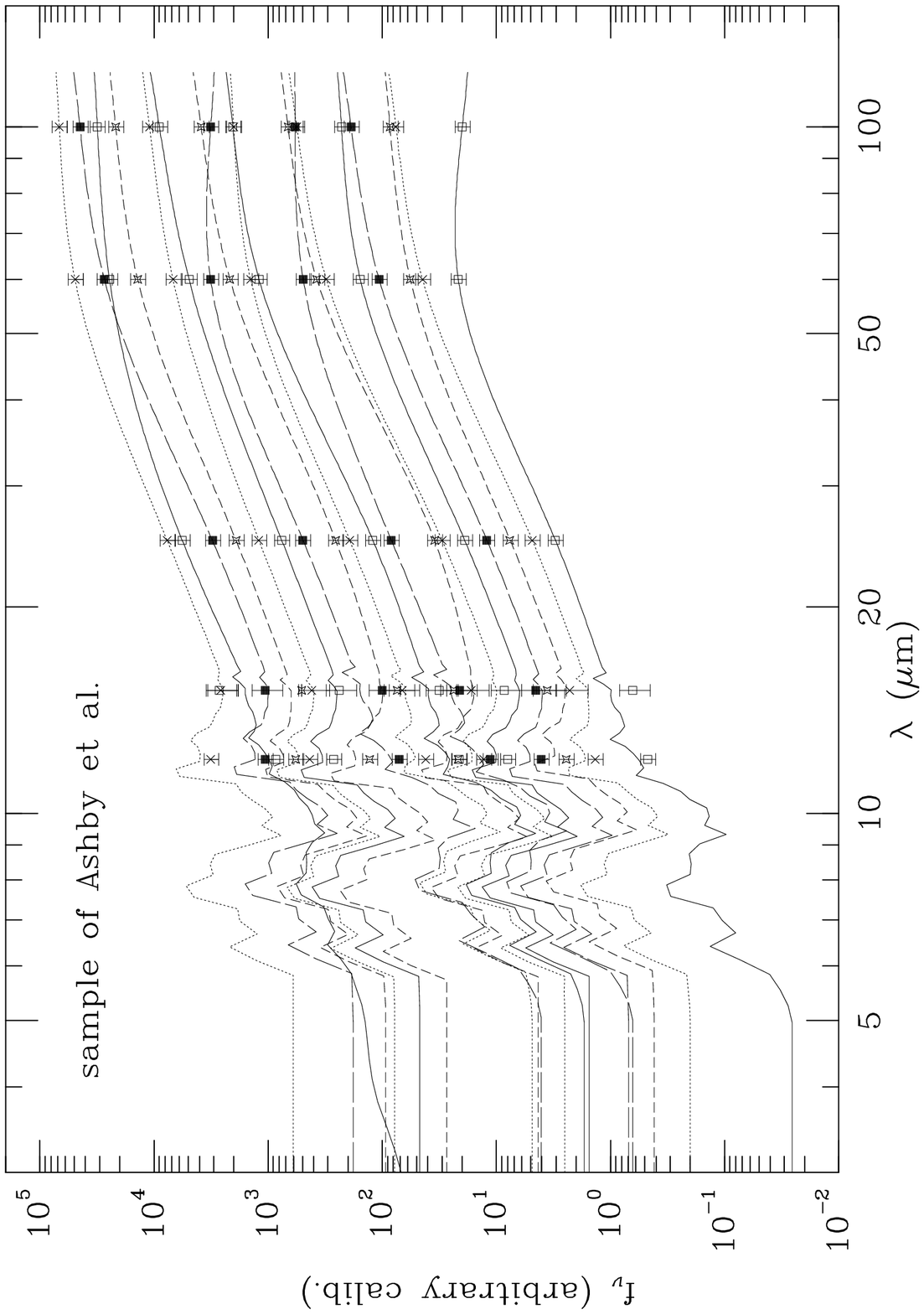]{
Model spectra of galaxies in the sample of Ashby{\it et al.},
overlaid by observed flux densities (IRAS: 12~$\mu$m, 25~$\mu$m, 60~$\mu$m
and 100~$\mu$m; ISOPHT: 16~$\mu$m). 
\label{fig:modspect}
}

\figcaption[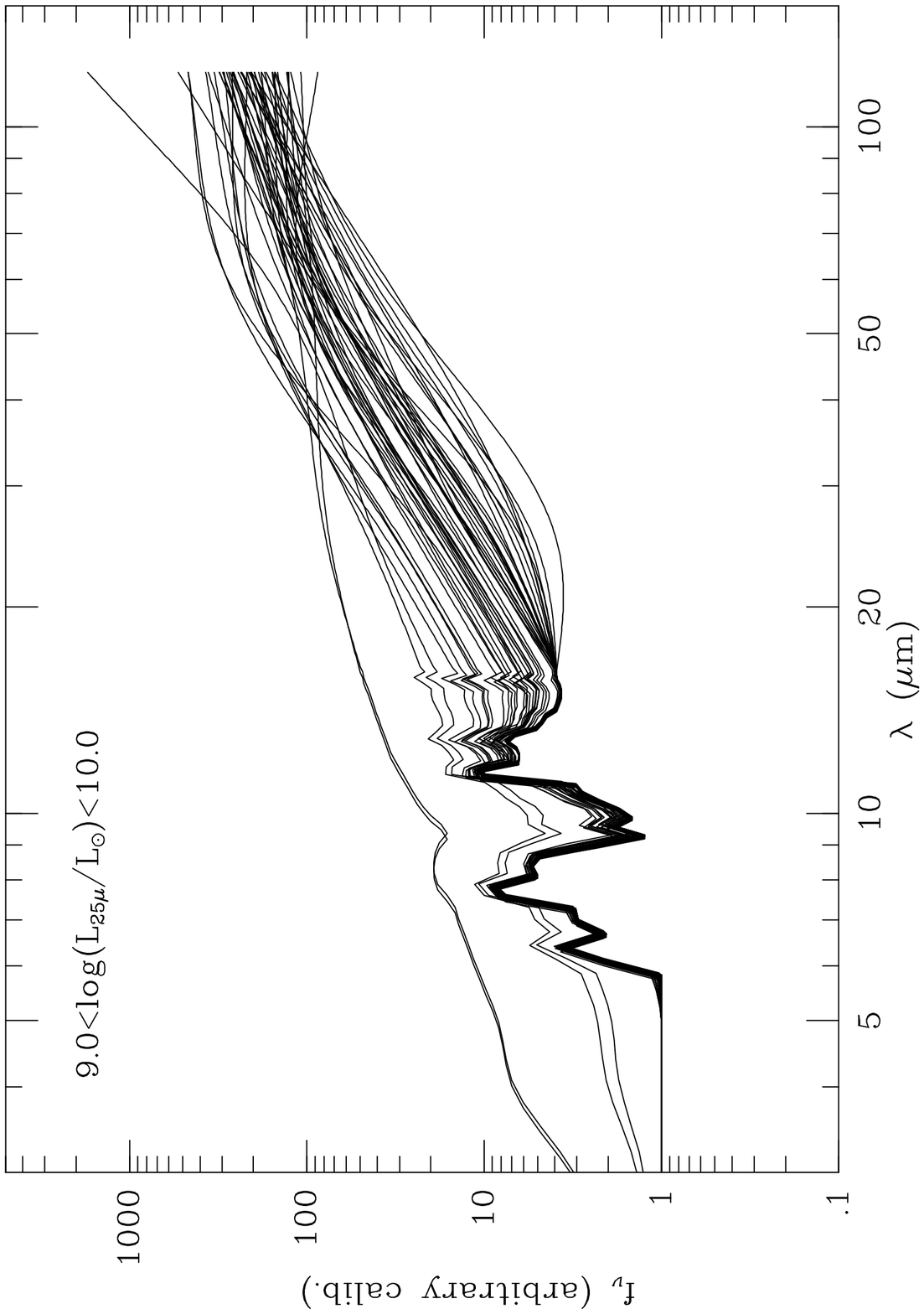]{
Model spectra of individual galaxies in the luminosity range
$9< log(L_{25\mu}/L_\sun) < 10$.
\label{fig:sed9-10}
}

\figcaption[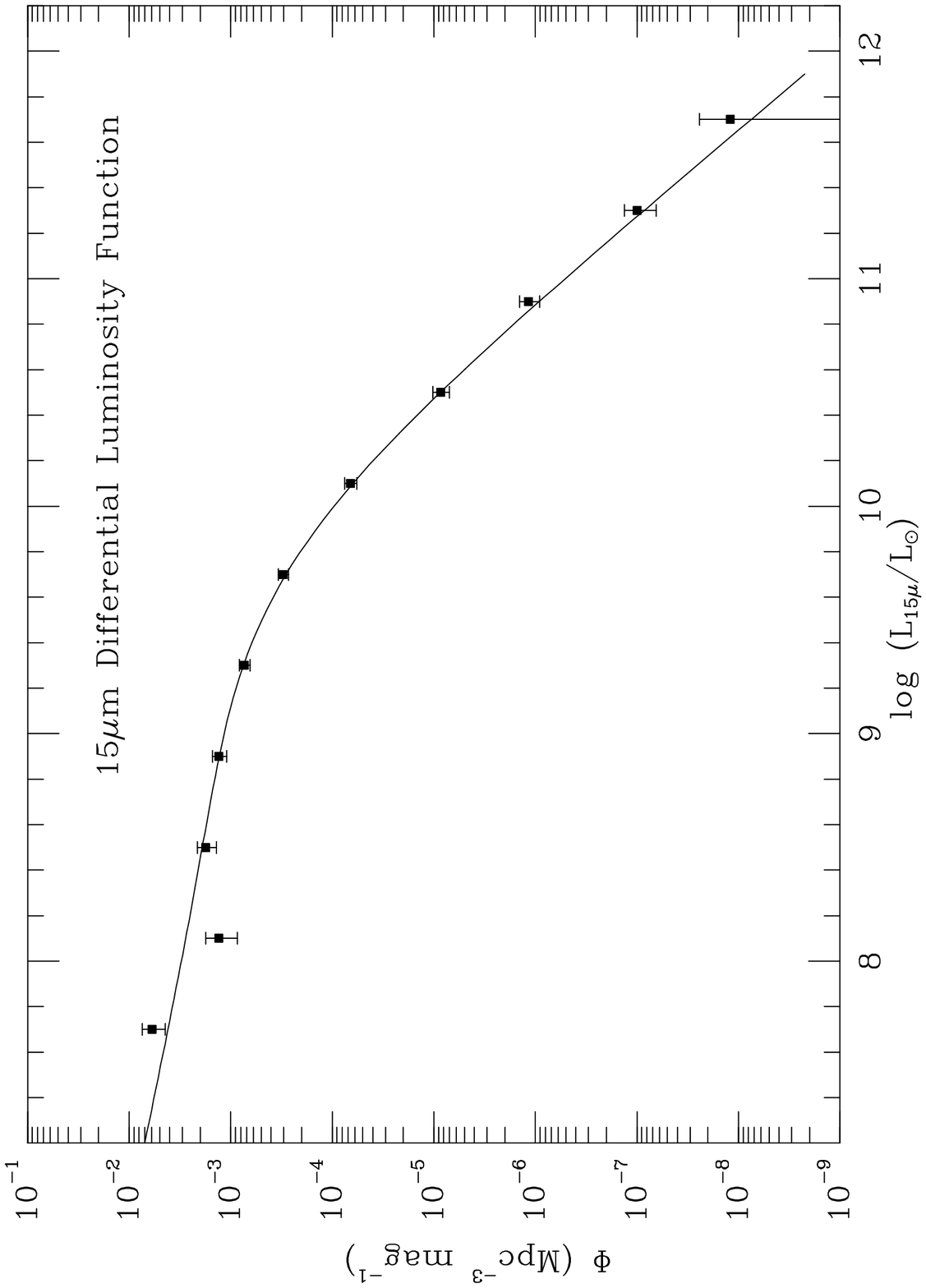]{
The 15~$\mu$m local differential luminosity function (Table 1).
The solid line is the smooth fit specified by the parameters given in
Table 2.
\label{fig:alcurv}
}

\figcaption[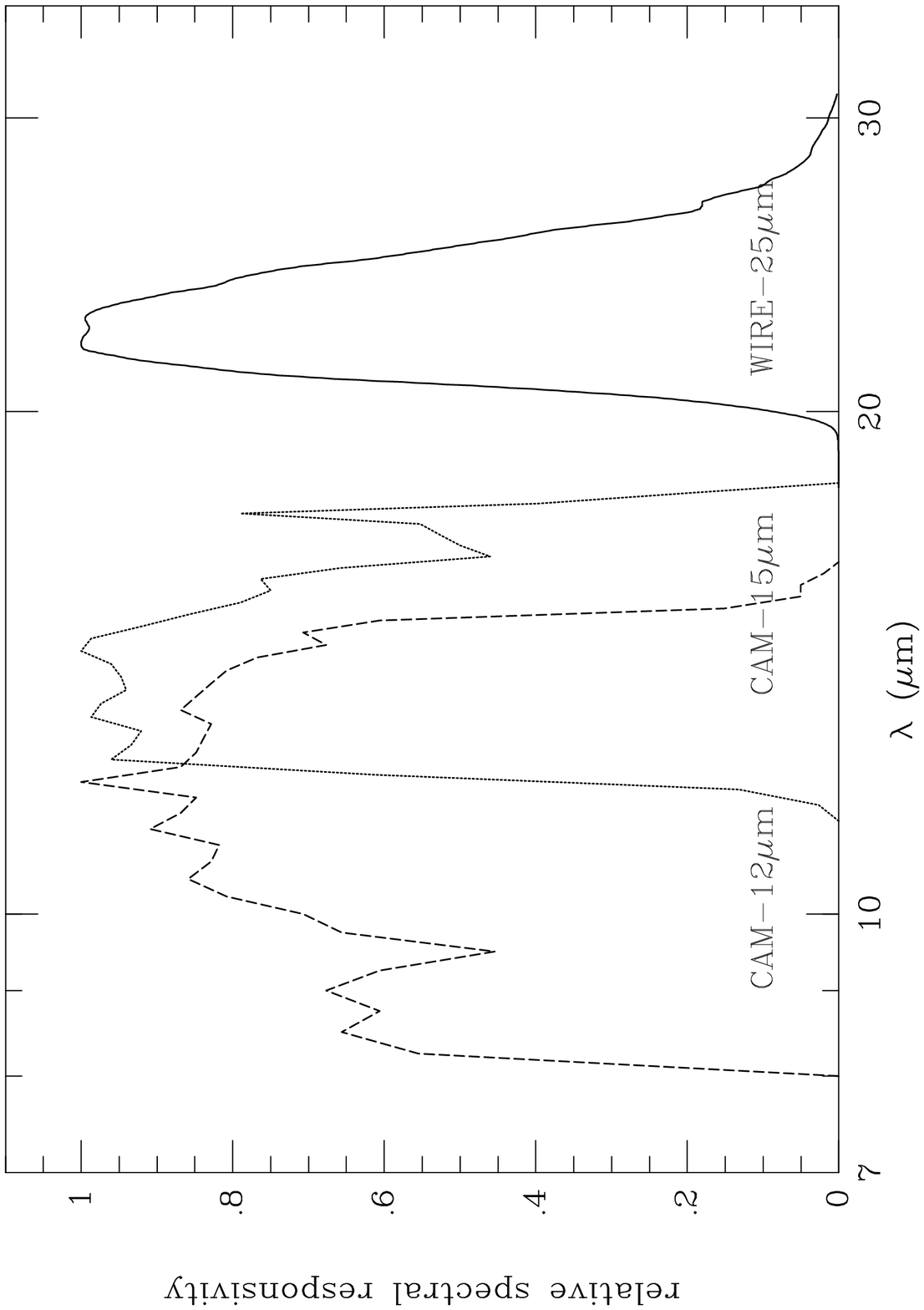]{
The bandpasses of the ISOCAM 12~$\mu$m and 15~$\mu$m filters and of the
WIRE 25~$\mu$m camera.
\label{fig:bandpass}
}

\figcaption[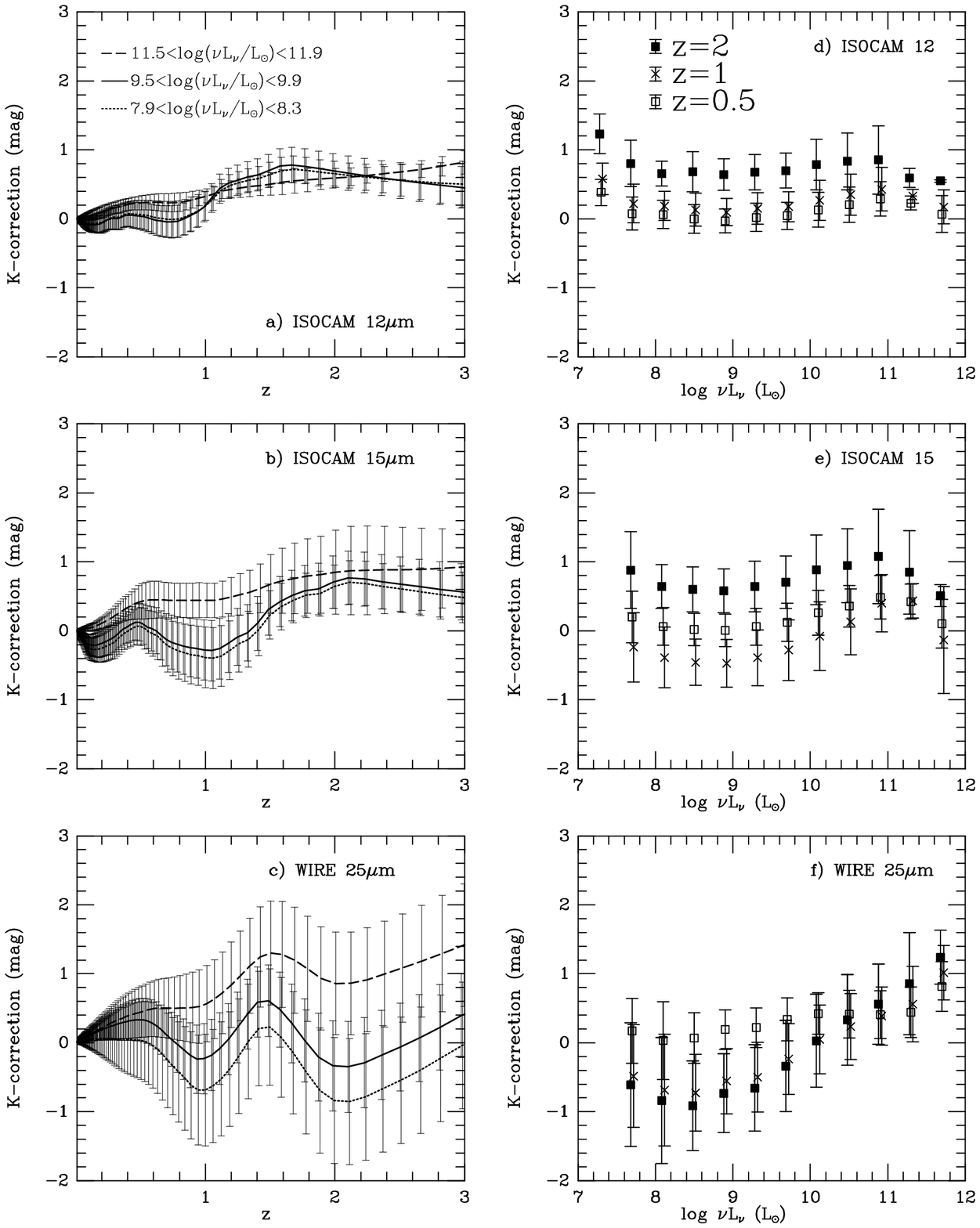]{
{\bf Panel a.} 
The mean $K_0$ and the dispersion $\sigma_K$ (shown by the error bars)
of K-corrections of the ISOCAM 12~$\mu$m flux as functions of
redshift $z$ for galaxies in three different luminosity bins.
{\bf Panel b.} Same as Panel a. For the ISOCAM 15~$\mu$m band.
{\bf Panel c.} Same as Panel a. For the WIRE 25~$\mu$m band.
{\bf Panel d.} Dependences of $K_0$ and $\sigma_K$ on the luminosity in the ISOCAM
12~$\mu$m band, plotted for three given redshift: $z=0.5$, $z=1$ and $z=2$. 
{\bf Panel e.} Same as Panel d. For the ISOCAM 15~$\mu$m band.
{\bf Panel f.} Same as Panel d. For the WIRE 25~$\mu$m band.
\label{fig:kcorr}
}

\figcaption[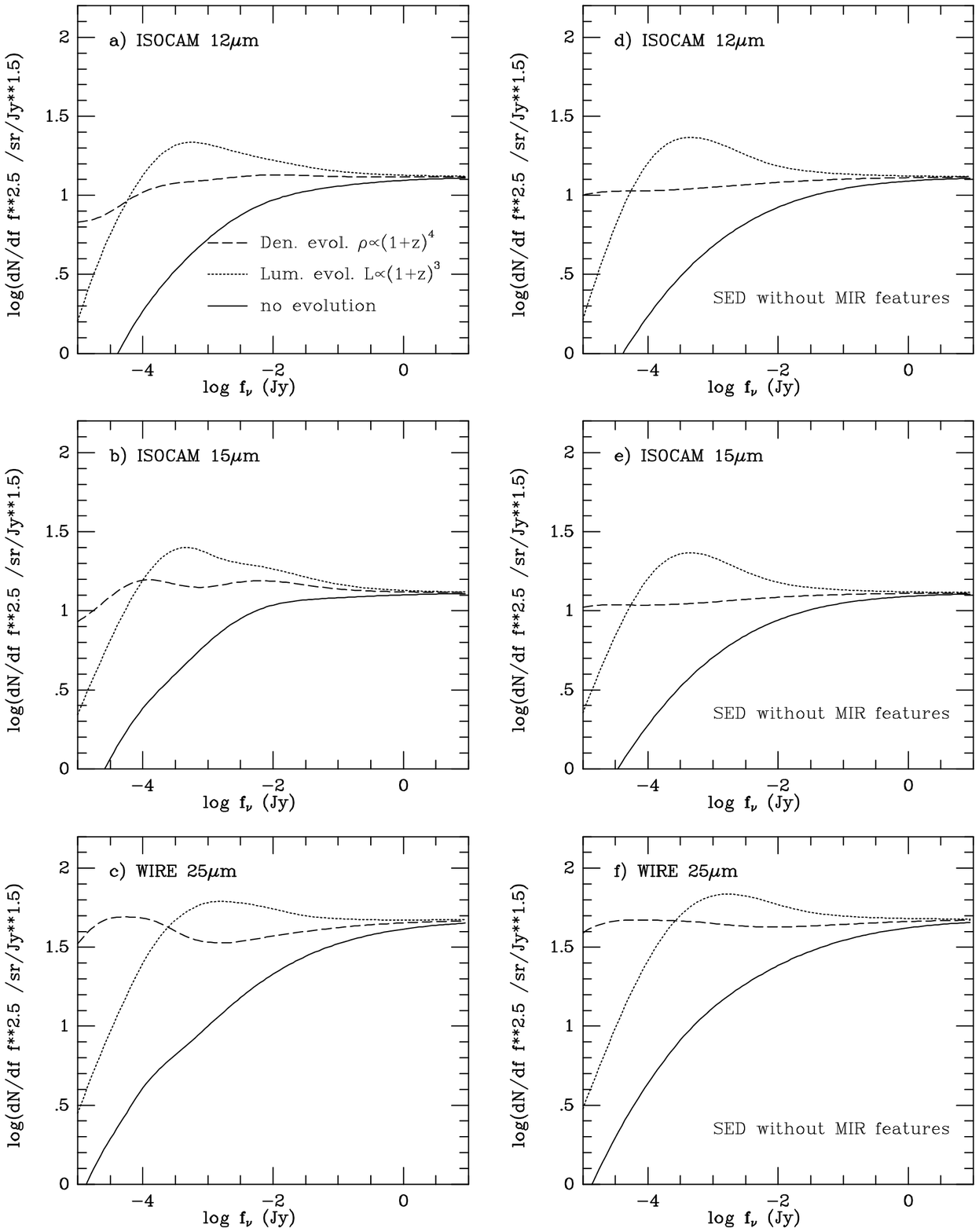]{
{\bf Panel a.} 
Euclidean normalized differential counts in the ISOCAM 12~$\mu$m band,
predicted by three different evolution models.
{\bf Panel b.} Same as Panel a. For the ISOCAM 15~$\mu$m band.
{\bf Panel c.} Same as Panel a. For the WIRE 25~$\mu$m band.
{\bf Panel d.} 
Euclidean normalized differential counts in the ISOCAM 12~$\mu$m band,
predicted by evolution models otherwise identical to those in Panel a
except that K-corrections are estimated from 
model spectra of galaxies without the MIR features.
{\bf Panel e.} Same as Panel d. For the ISOCAM 15~$\mu$m band.
{\bf Panel f.} Same as Panel d. For the WIRE 25~$\mu$m band.
\label{fig:fcndis}
}

\figcaption[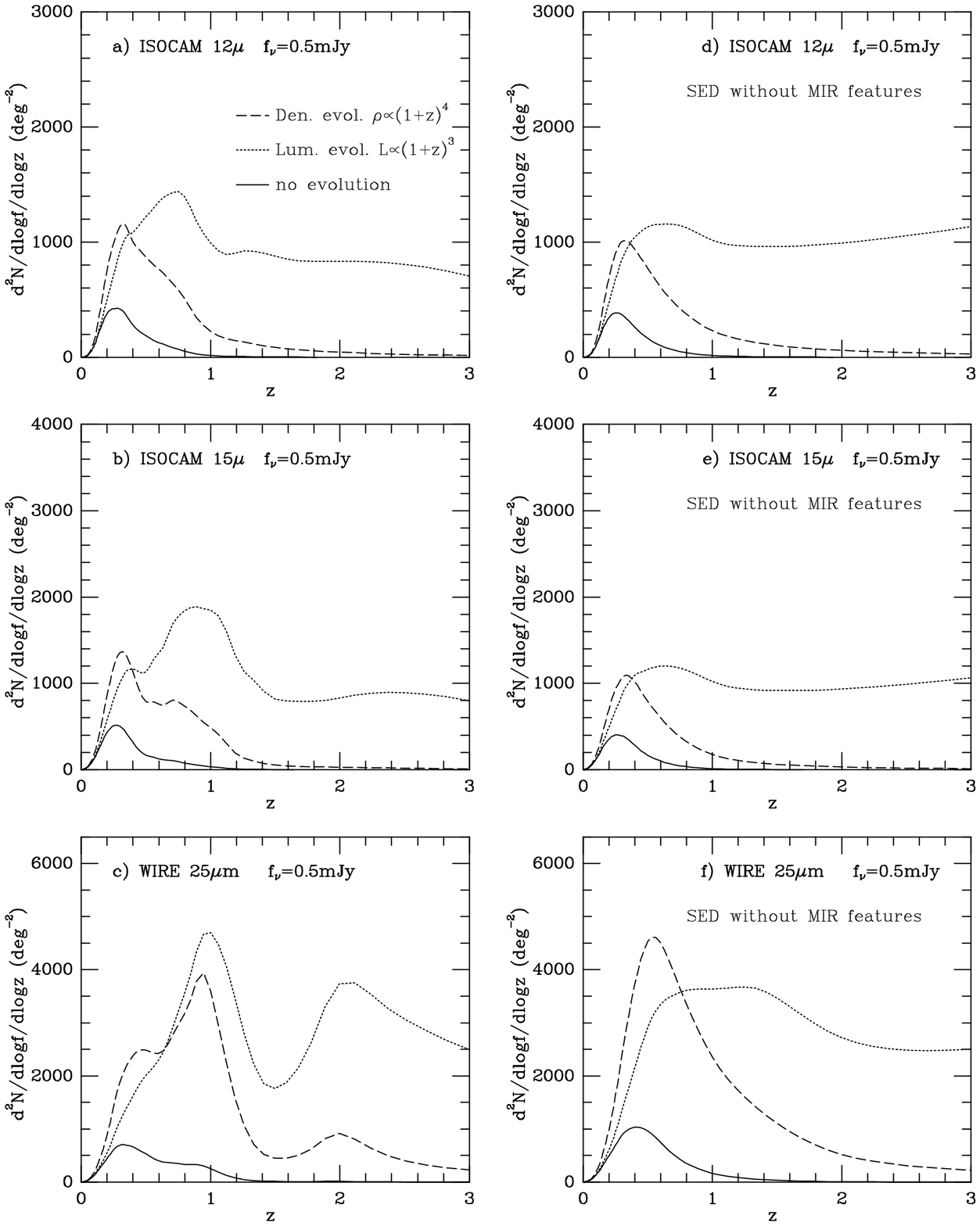]{
{\bf Panel a.} 
Redshift distributions
of sources with $f_\nu = 0.5$~mJy in the ISOCAM 12~$\mu$m band
predicted by the three evolution models
(note that the distributions are
for galaxies at this flux level, rather than for galaxies in
samples brighter than this flux).
{\bf Panel b.} Same as Panel a. For the ISOCAM 15~$\mu$m band.
{\bf Panel c.} Same as Panel a. For the WIRE 25~$\mu$m band.
{\bf Panel d.} 
Redshift distributions
of sources with $f_\nu = 0.5$~mJy in the ISOCAM 12~$\mu$m band
predicted by the evolution models
otherwise identical to those in Panel a
except that K-corrections are estimated from 
model spectra of galaxies without the MIR features.
{\bf Panel e.} Same as Panel d. For the ISOCAM 15~$\mu$m band.
{\bf Panel f.} Same as Panel d. For the WIRE 25~$\mu$m band.
\label{fig:rds}
}

\end{document}